\documentclass[twocolumn]{aastex6}%
\pdfoutput=1
\usepackage{natbib}
\usepackage{amsmath}
\usepackage{multirow}
\usepackage{graphicx}
\usepackage{color}
\usepackage{threeparttable}
\usepackage{enumitem}
\usepackage{lipsum}

\usepackage{lipsum}

\newcommand\blfootnote[1]{%
  \begingroup
  \renewcommand\thefootnote{}\footnote{#1}%
  \addtocounter{footnote}{-1}%
  \endgroup
}

\def\maxij1820{MAXI~J1820+070}
\def\gx339{GX~339--4}
\def\exo1846{EXO~1846--031}
\def\at2019wey{AT2019wey}
\def\igr{IGR J17091--3624}
\def\grs{GRS~1915+105}

\def\h1743{H1743--322}

\begin{document}
\title{Highly-coherent quasi-periodic oscillations in the `heartbeat' black hole X-ray binary \igr}
\author{\vspace{-1.5cm}Jingyi~Wang\altaffilmark{1},
Erin~Kara\altaffilmark{1}, 
Jeroen~Homan\altaffilmark{2},
James~F.~Steiner\altaffilmark{3},
Diego~Altamirano\altaffilmark{4},
Tomaso~Belloni\altaffilmark{5}$^{,*}$,
Michiel van der Klis\altaffilmark{6}, 
Adam~Ingram\altaffilmark{7}, 
Javier~A.~Garc{\'\i}a \altaffilmark{8,9}, 
Guglielmo~Mastroserio\altaffilmark{10}, 
Riley~Connors\altaffilmark{11}, 
Matteo~Lucchini\altaffilmark{6},
Thomas~Dauser\altaffilmark{12}, 
Joseph~Neilsen\altaffilmark{11}, 
Collin~Lewin\altaffilmark{1},
Ron~A.~Remillard\altaffilmark{1},
}
\affil{
\altaffilmark{1}MIT Kavli Institute for Astrophysics and Space
Research, MIT, 70 Vassar Street, Cambridge, MA 02139, USA\\
\altaffilmark{2}Eureka Scientiﬁc, Inc., 2452 Delmer Street, Oakland, CA 94602, USA\\
\altaffilmark{3}Harvard-Smithsonian Center for Astrophysics, 60 Garden St., Cambridge, MA 02138, USA \\
\altaffilmark{4}School of Physics and Astronomy, University of Southampton, Highfield, Southampton, SO17 1BJ\\
\altaffilmark{5}INAF-Osservatorio Astronomico di Brera, via E. Bianchi 46, I-23807 Merate, Italy\\
\altaffilmark{6}Astronomical Institute, Anton Pannekoek, University of Amsterdam, Science Park 904, NL-1098 XH Amsterdam, Netherlands\\
\altaffilmark{7}School of Mathematics, Statistics and Physics, Newcastle University, Herschel Building, Newcastle upon Tyne, NE1 7RU, UK\\
\altaffilmark{8}X‐ray Astrophysics Laboratory, NASA Goddard Space Flight Center, Greenbelt, MD 20771\\ 
\altaffilmark{9}Cahill Center for Astronomy and Astrophysics, California Institute of Technology, Pasadena, CA 91125, USA\\  
\altaffilmark{10}INAF-Osservatorio Astronomico di Cagliari, via della Scienza 5, I-09047 Selargius (CA), Italy\\
\altaffilmark{11}Villanova University, Department of Physics, Villanova, PA 19085, USA\\
\altaffilmark{12}Remeis Observatory \& ECAP, Universit\"{a}t
Erlangen-N\"{u}rnberg, 96049 Bamberg, Germany\\
}

\begin{abstract}

\igr\ is a black hole X-ray binary (BHXB), often referred to as the `twin' of \grs\ because it is the only other known BHXB {\color{black}that can} show exotic `heartbeat'-like variability {\color{black}that is highly structured and repeated}. Here we report on observations of \igr\ from its 2022 outburst, where we detect an unusually coherent quasi-periodic oscillation (QPO) when the {\color{black}broadband} variability is low (total fractional rms $\lesssim6\%$) and the spectrum is dominated by the accretion disk. 
{\color{black}Such spectral and variability behavior is characteristic of the soft state of typical BHXBs (i.e. those that do not show heartbeats), but we also find that this QPO is strongest when there is some exotic heartbeat-like variability (so-called Class V variability).}
This QPO is detected at frequencies between 5 and 8~Hz and has Q-factors (defined as the QPO frequency divided by the width) $\gtrsim50$, making it {\color{black}one of the} most highly coherent low-frequency QPO ever seen in a BHXB. 
The extremely high Q factor makes this QPO distinct from typical low-frequency QPOs that are conventionally classified into Type-A/B/C QPOs. Instead, we find evidence that archival observations of \grs\ also showed a similarly high-coherence QPO in the same frequency range, suggesting that this unusually coherent and strong QPO may be unique to {\color{black}BHXBs that can exhibit `heartbeat'-like variability.}.

\end{abstract}

\section{Introduction}\label{intro}
BHXBs provide us with opportunities to study different accretion states of accreting black holes in a single source (see {\color{black}e.g., \citealp{1997ApJ...479..926M,remillard2006x,2011BASI...39..409B}, and \citealp{2022arXiv220614410K} for a recent review). {\color{black}In a typical outburst, the BHXBs} rise from {\color{black}quiescence through a} \textit{hard state} where the X-ray emission is dominated by emission from the `corona' ({\color{black}the} hot plasma with temperature on the order of $100$~keV). Then, {\color{black}the BHXBs} make a transition in days {\color{black}to weeks}, through what is known as the \textit{intermediate state}, into the \textit{soft state} where the disk emission dominates. Finally, they return to {\color{black}the} hard state and then {\color{black}recede again into} quiescence. }\blfootnote{$^*$We dedicate this paper to the late Tomaso Belloni, who contributed significantly to this paper before his untimely passing on 26 August 2023. Tomaso was a pioneer in the study of X-ray timing since his early days working on EXOSAT, and in particular, awakened the community to the beautiful puzzle that is GRS 1915. In this work, on GRS 1915's `little sister', IGR J17091, we build upon the legacy of a trailblazer in our field. We will miss him for his energy, his insights, his humor and his unwavering passion for science. Ad astra, Tomaso.}

\begin{figure}
\centering
\includegraphics[width=1.\linewidth]{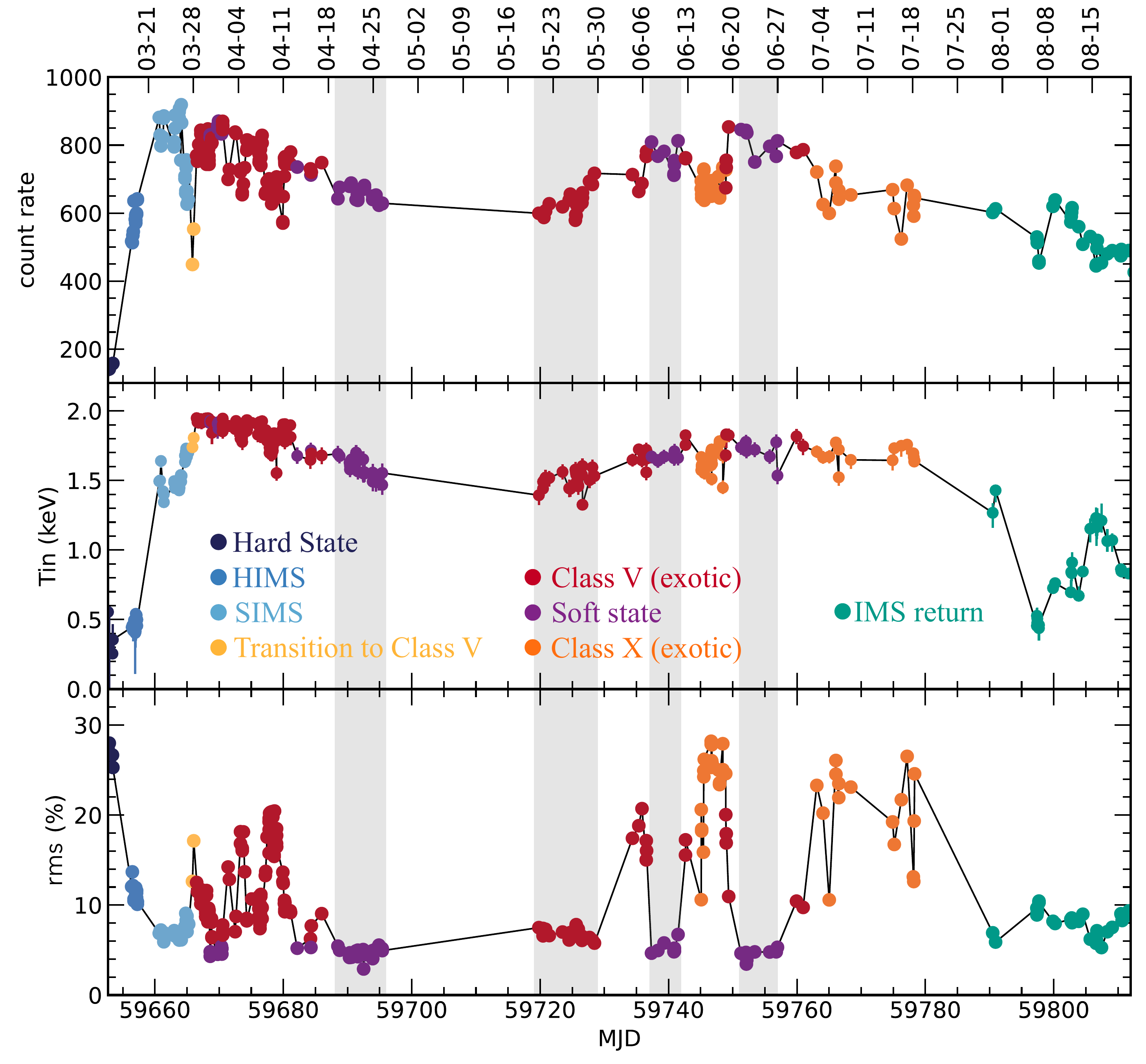}
\caption{The time evolution of NICER count rate (0.3--12~keV, normalized for 52~FPMs), the fitted disk temperature with a baseline model (W24a), and the fractional rms (0.01--10~Hz in 1--10~keV). There are 305 data points, each representing a 500~s NICER segment. 
The gray shaded regions indicate when the highly-coherent QPO was detected (Epochs 1--10 in Table~\ref{tab}).
Besides MJD, the calendar dates are shown on the top x-axis. {\color{black}Classes V and X are termed `exotic' because exotic `heartbeat'-like variability (structured and repeated) is present in the lightcurves.} See {\color{black}Section~\ref{obs}} and W24a for more details on the state identifications.}
\label{fig:mjd}
\end{figure}


{\color{black}In the lightcurves of BHXBs, we observe various types of low-frequency quasi periodic oscillations \citep[LFQPOs; see the reviews][and references therein]{2016ASSL..440...61B,ingram2019review}.} The LFQPOs in BHXBs are usually categorized with an {\color{black}A/B/C} classification scheme (see e.g., \citealp{1999ApJ...526L..33W,2002ApJ...564..962R,2005ApJ...629..403C,2011MNRAS.418.2292M}). {\color{black}The classification is based on the properties of the QPO including e.g., its frequency, strength, and Q factor (frequency divided by the full width at half maximum, FWHM), the type of the underlying noise component, and the spectral state in which {\color{black}the QPO} is detected.}

\begin{figure}
\centering
\includegraphics[width=1.\linewidth]{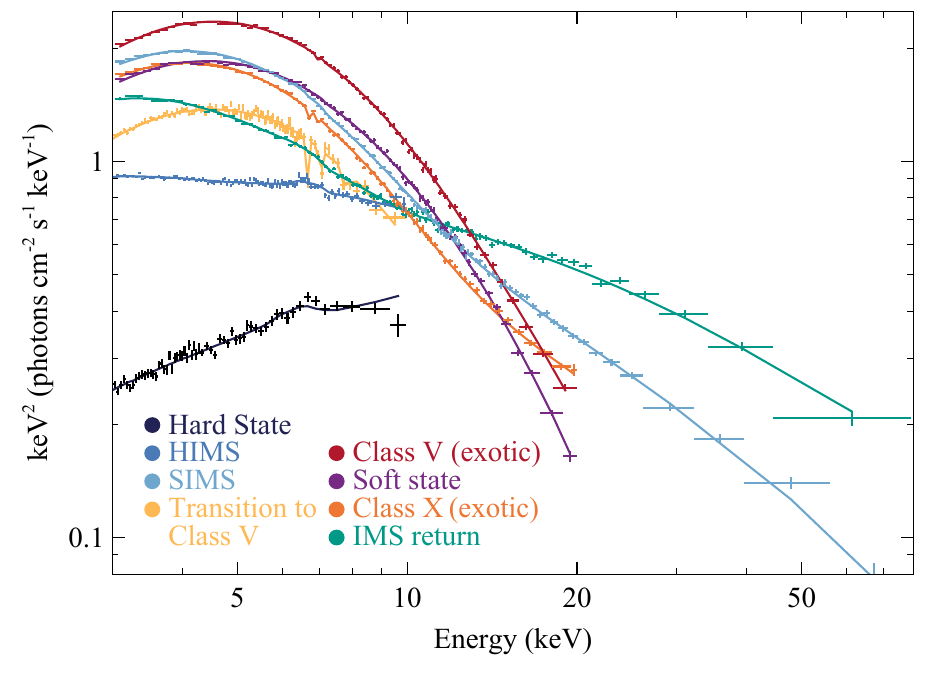}
\caption{{\color{black}The unfolded spectra using a model where reflection and absorption lines are included (see W24a for more details). For each state, when a simultaneous NuSTAR spectrum is available, we show that for coverage at high energies; in the \textit{Hard state}, \textit{HIMS}, and \textit{Transition to Class V}, the NICER spectra are shown. The highly coherent QPO detected are in the \textit{Soft State} and \textit{Class V}, which are both disk-dominated, consistent with a high disk temperature as illustrated in Fig.~\ref{fig:mjd}.} }
\label{fig:spec}
\end{figure}

\begin{itemize}

\item Type-C QPOs are seen commonly in the hard state and hard-intermediate state (HIMS). The QPO frequency ranges between a few mHz to $\sim10$~Hz. They are strong ($\lesssim20\%$ rms), narrow ($Q\gtrsim8$), and accompanied by a flat-top noise. In addition to the fundamental (usually defined as the component with the highest rms amplitude), there are {\color{black}often} second and even third harmonics and the sub-harmonic (half of the frequency). The frequency of Type-C QPO was found to be correlated with the spectral properties of both the disk and the corona, {\color{black}albeit with some spread. For instance, correlations have been seen with} e.g., the disk flux \citep{1999ApJ...513L..37M,2000ApJ...531..537S,remillard2006x}, the photon index of the {\color{black}coronal spectrum} \citep{2003A&A...397..729V,2022MNRAS.513.4196G}{\color{black}, and the coronal electron temperature \citep{2022NatAs...6..577M}.} Compared with {\color{black}other} timing properties, the QPO frequency is also correlated with the low-frequency break of the flat-top noise \citep{1999ApJ...514..939W,1999ApJ...520..262P}, and is anticorrelated with the total fractional rms \citep{2011MNRAS.418.2292M}. There is evidence that Type-C QPOs have inclination dependency of their amplitudes \citep{2015MNRAS.447.2059M} and the time lags at the QPO frequencies \citep{2017MNRAS.464.2643V}. This may suggest that they are produced by some geometrical effects involving Lense-Thirring precession {\color{black}(e.g, the Relativistic Precession Model, \citealp{1998ApJ...492L..59S}; the solid-body Lense-Thirring precessing model, \citealp{2009MNRAS.397L.101I}; precession of the base of a jet,  \citealp{2021NatAs...5...94M}).} However, recently, \citet{2021ApJ...906..106M} show that not all models of the inner accretion flow can produce precession. {\color{black}Besides geometrical models, there are also models involving intrinsic luminosity changes of the system including e.g., accretion ejection instability model \citep{1999A&A...349.1003T}, and variable Comptonization \citep{2022MNRAS.515.2099B}.}

\item Type-B QPOs are seen in soft-intermediate state (SIMS), and they are narrow ($Q\gtrsim6$) but weaker compared to Type-C ($\lesssim5\%$ rms), found usually at 5--6~Hz and sometimes 1--3 Hz, and appear on top of weak red noise (few percent rms). The QPO frequency is not anticorrelated with the total fractional rms as observed for Type-C QPOs \citep{2011MNRAS.418.2292M}. The appearance of the Type-B QPO has been suggested to be linked to discrete jet ejections \citep{2004MNRAS.355.1105F,2016MNRAS.460.2796S,kylafis2020,2021MNRAS.501.3173G,liu2022,2023MNRAS.525..854M}. {\color{black}The best example is a Type-B QPO in \maxij1820 that appeared at a time consistent with that of a jet ejection event \citep{2020ApJ...891L..29H,2021MNRAS.505.3393W}.}

\item Type-A QPOs also normally appear in the SIMS. They are the least common type as there were only $\sim$10 detections in the entire RXTE archive. The QPO frequency is between 6 and 8~Hz. They are weak (a few percent rms), broad ($Q\lesssim3$), and they are accompanied by very weak red noise \citep{2001ApJS..132..377H,2011MNRAS.418.2292M,2015MNRAS.447.2059M}. As they are rare, weak, and broad, the phenomenology and nature of Type-A QPOs are the least understood. 

\end{itemize}


\igr\ and \grs\ are extraordinary BHXBs because they are the only two {\color{black}known} BHXBs that, in addition to LFQPOs, {\color{black}exhibit} a variety of \textit{exotic variability} {\color{black}classes}, usually consisting of flares and drops that are highly structured and have high amplitudes (e.g., \citealp{2000A&A...355..271B,altamirano2011faint,2017MNRAS.468.4748C}). 
{\color{black}Because of a famous `heartbeat' class when the lightcurve resembles an electrocardiogram (\textit{Class IV} in \igr\ and \textit{Class $\rho$} in \grs), in this work, we refer to variabilities that are structured and repeated as `exotic' or `heartbeat-like'.}
{\color{black}We observe LFQPOs in the X-ray light curves of these two BHXBs, mostly in the hard state and intermediate state without exotic variabilities \citep{1997ApJ...482..993M,2000ApJ...541..883R,altamirano2011faint,2014ApJ...783..141P,wang24a}. LFQPOs are sometimes also detected in the states showing exotic variabilitites, and the QPO changes in frequency and even type along with the exotic variabilities \citep{1999ApJ...513L..37M,1999ApJ...527..321M,2002A&A...386..271R,2004ARA&A..42..317F,2008MNRAS.383.1089S,2011ApJ...737...69N,2017MNRAS.468.4748C}. } 

Besides the LFQPOs, there are also high-frequency QPOs (HFQPOs) in BHXBs but these are detected {\color{black}less frequently}. In the RXTE archive, HFQPOs were only detected in a handful of BHXBs, and some detections are statistically marginal (see \citealp{2012MNRAS.426.1701B} and references therein). The QPO frequency is usually a few hundred Hz, while it is at only $\sim67$~Hz in both \grs\footnote{{\color{black}We note that \citet{mobe2023} recently suggested that the 67 Hz QPO in GRS 1915+105 may in fact be a Type-C QPO, rather than a HFQPO.}} and \igr\ \citep{1997ApJ...482..993M,2012ApJ...747L...4A}. The Q factor is in the range of 5--30, with an amplitude of 0.5--6\%. HFQPOs sometimes show up as two peaks with a frequency ratio close to an  integer ratio of 3:2, but with rare exception, the pair are not simultaneous or appear in different energy bands \citep{2001ApJ...552L..49S,2002ApJ...580.1030R,remillard2006x}.


In \citet{wang24a} (hereafter W24a), we presented the spectral-timing analysis of \igr\ in its 2022 outburst using NICER, NuSTAR, and Chandra data. We found that, as in typical BHXBs, the outburst began in the hard state, then the intermediate state, but then transitioned to a soft state where we identified two types of heartbeat-like variability (Class V and a new Class X; see also Fig.~\ref{fig:mjd}). We observed Type-C QPOs in the hard state and HIMS, Type-B QPOs in the SIMS, as in normal BHXBs, and a highly-coherent QPO in the soft state {\color{black}and Class V}. In this paper, we focus on the highly-coherent QPO, including its evolution and properties in Section~\ref{sect:results}, and discuss its nature in Section~\ref{sect:discussion}.

\begin{figure}
\centering
\includegraphics[width=1.\linewidth]{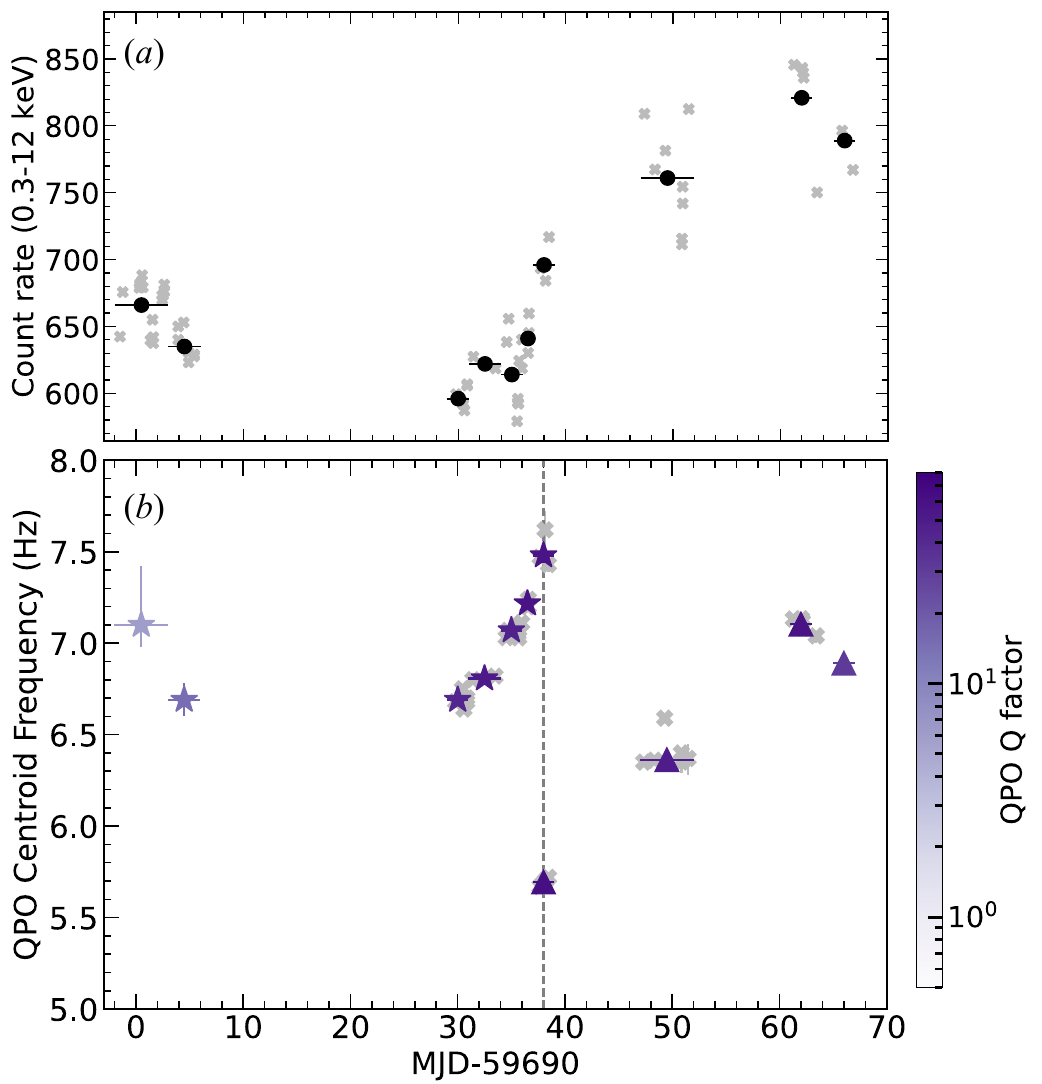}
\caption{The time evolution of (a) the count rate in 0.3--12~keV (normalized to 52~FPMs), and (b) the QPO centroid frequency. The {\color{black}purple} data points are measurements from the 10 data epochs corresponding to the PSDs in Fig.~\ref{fig:new_qpo}, and the gray underlying data points are single-segment measurements. {\color{black}The dashed line indicates Epoch~7 when two QPO components are detected.} The star and triangle markers represent the two QPO components respectively, and the color scale represents the Q factor of the QPO. {\color{black}In (b), all the QPOs in purple (from the 10 data epochs) are also significantly detected above $3\sigma$ in single-segment PSDs, except for the QPOs in Epochs~1, 2, which are also the least significant QPOs (below $6\sigma$) as shown in Table~\ref{tab}.}
\label{fig:qpo_evolution_mjd}}
\end{figure}

\begin{figure*}
\centering
\includegraphics[width=1.\linewidth]{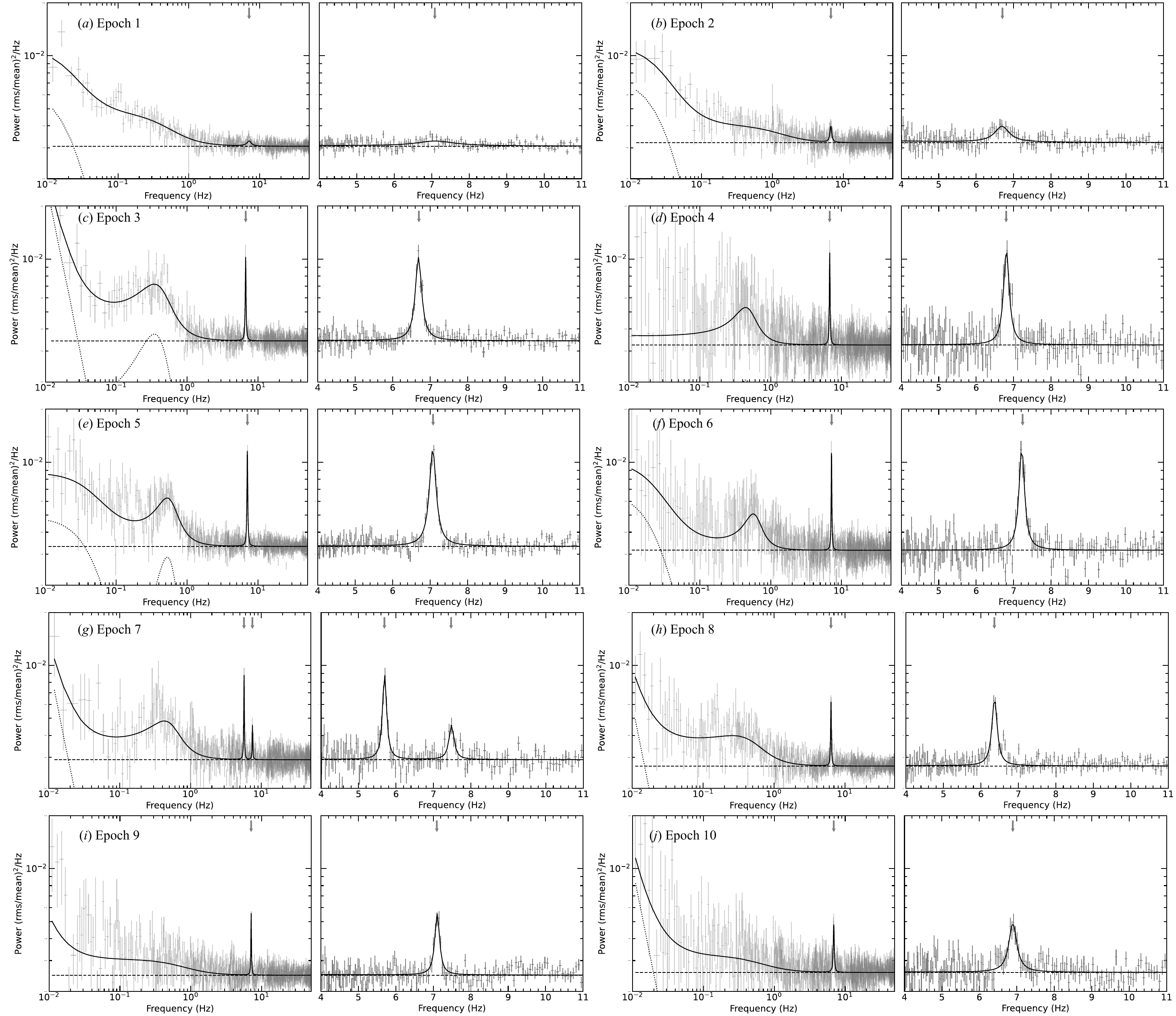}
\caption{The PSDs that show the highly-coherent QPOs. The PSDs are raw, as the Poisson noise are not subtracted. {\color{black}For each epoch, a zoom-in of the PSD at the QPO frequency band is shown on the right. The solid line shows the best-fit with multiple Lorentzians, the dashed line shows the Poisson noise level, and the dotted lines show the other noise components. The Lorentzian components for the QPOs are not shown for clarity. 
The gray arrows indicate the fitted QPO centroid frequencies (corresponding to the values in Table~\ref{tab}). 
{\color{black}Representative corresponding lightcurves can be found in W24a.}}
}
\label{fig:new_qpo}
\end{figure*}


\section{Observations and data reduction}\label{obs}
After its last outburst in 2016, \igr\ entered a new outburst in March 2022 \citep{miller2022nicer}. When this outburst began, we triggered our NICER and NuSTAR GO Program (PI: J. Wang). We analyzed all 136 NICER observations taken at a near-daily cadence from 2022 March 27 to Aug 21 in W24a. Readers are referred to W24a for data reduction and state classification. {\color{black}In summary, we identify the different states by the broadband spectral shape, the power spectral densities (PSDs), and the shape of the lightcurves. From the analysis of the broadband spectral shape (Figs.~\ref{fig:mjd} and \ref{fig:spec}) and PSDs, we conclude that \igr\ transitions between spectral and timing characteristics of typical BHXBs (the \textit{Hard State}, \textit{HIMS}, \textit{SIMS}, \textit{Soft State} and \textit{IMS Return}). Then, the shape of the lightcurves revealed that there was a transitional phase termed \textit{Transition to Class V}, and that sometimes in the soft state when the disk dominates over the coronal emission, instead of showing very little variability (as in most BHXBs), \igr\ can demonstrate exotic (structured and repeated) variability. Therefore, we also identified exotic variability classes \textit{Class V} and a new class that we labeled \textit{Class X}. To be self-contained, we included two figures modified from W24a. Fig.~\ref{fig:mjd} presents the time evolution of the count rate, fitted disk temperature, and fractional rms, with the gray-shaded regions highlighting the data in which the highly coherent QPO was detected. We found the highly coherent QPO was detected in the \textit{Soft State} and \textit{Class V} when the disk temperature was high (1.5--2~keV; middle panel), the coronal spectrum was soft (photon index $\sim2.7$; see W24a), and the {\color{black}broadband} variability was weak (low rms $\lesssim6\%$ {\color{black}in 0.01--10~Hz}; bottom panel). 
Fig.~\ref{fig:spec} shows the unfolded spectra using a model where disk and coronal emission, reflection, and absorption lines are included (see W24a for more details). We see that \igr\ transitioned from a hard and corona-dominated state to a soft and disk-dominated state. The highly coherent QPO was detected in disk-dominated states (\textit{Soft State} and \textit{Class V}).
}

{\color{black}In this paper,} we compute PSDs {\color{black}in 1--10~keV} using a segment length of 500~s and a time bin of 1~ms (Nyquist frequency of 500~Hz). {\color{black}This allows us to capture both the LFQPOs and the exotic variability in a single PSD. However, due to our 500~s segment length requirement, several short ObsIDs with significant LFQPOs were excluded. We therefore also produced PSDs from segments with a length of 64\,s. The results from these shorter-segment PSDs are presented in Appendix A.} We use the `rms-squared' normalization and the PSDs are binned geometrically in frequency, i.e., from frequency $\nu$ to $(1+f)\nu$, where $f$ is called the f factor (see {\color{black}e.g.,} Section~2.2 in \citealt{uttley2014x} for more details). {\color{black}For single-segment PSDs, we use an f factor of 0.01. For PSDs with multiple segments, from 0.002 to 500~Hz, we choose 10 frequency ranges in logarithmic space where the f factor decreases logarithmically from 0.8 to 0.0008. This results in an f factor $\sim0.008$ in the frequency range where the highly-coherent QPOs are detected (5--8~Hz). We note that for single-segment PSDs, the errors on the PSD approach Gaussian at frequencies $\gtrsim3$~Hz ($\gtrsim20$~Foutier frequencies per frequency bin). When combining $\sim10$~segments, the PSD errors approach Gaussian at frequencies as low as $\sim0.01$~Hz, and these are the PSDs that the conclusions are ultimately made from.} We fit the raw (Poisson noise included) PSDs with a model including multiple Lorentzian components and a constant for the Poisson noise. Lorentzians are used to describe both the broadband noise and the QPOs.

All the uncertainties quoted in this paper are for a 90\% confidence range unless otherwise stated. We use XSPEC 12.12.1, and $\chi^2$ fit statistics for the power spectral fits. 

\section{Results} \label{sect:results}

\begin{figure*}
\centering
\includegraphics[width=1.\linewidth]{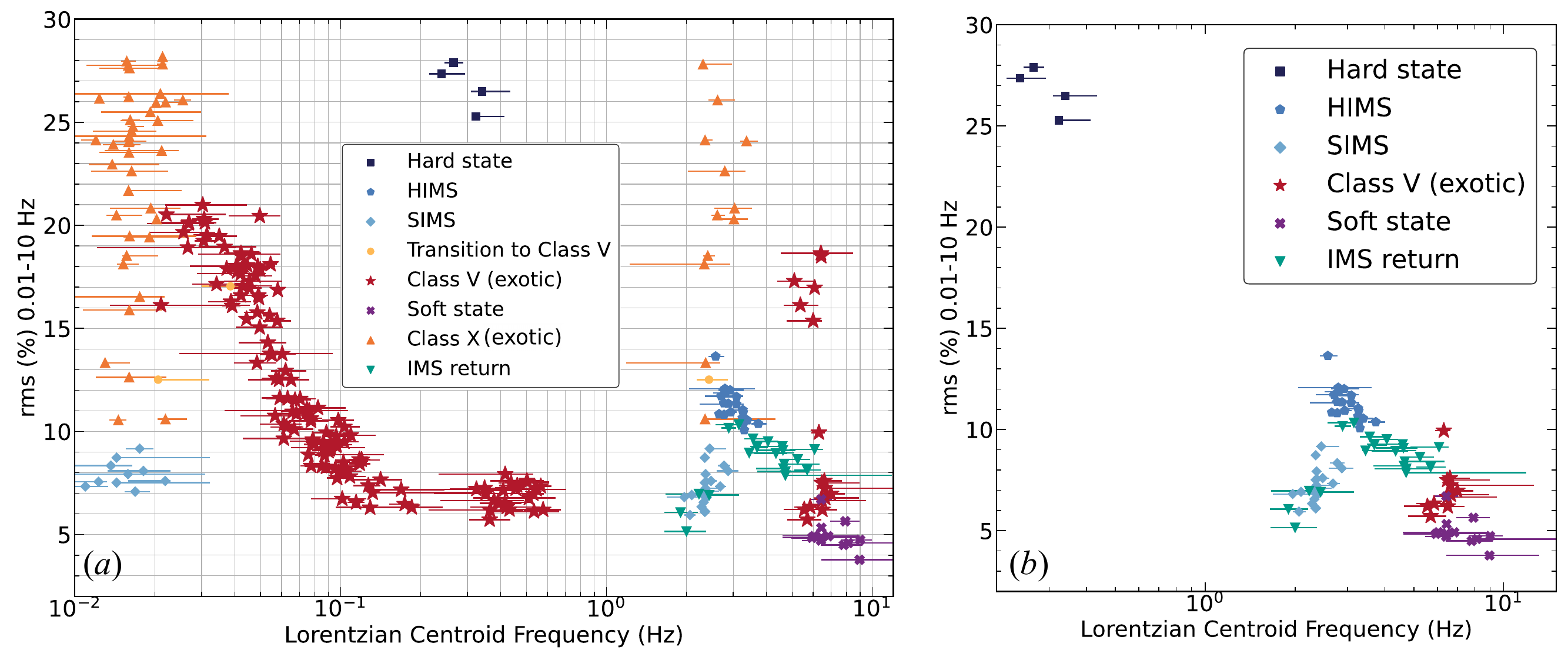}
\caption{{\color{black}The fitted Lorentzian centroid frequencies in the single-segment PSDs versus the fractional rms (0.01 to 10~Hz). In (a), we show the Lorentzian components for all the 8 states we identified in W24a; in (b), we only show the Lorentzian components representing the Type-C and Type-B QPOs in the hard state and intermediate state, and the highly coherent QPOs (in \textit{Class V} and \textit{Soft State}) that are the focus on this paper. The NICER energy band used is 1--10~keV. }}
\label{fig:rms_f0}
\end{figure*}

{\color{black}We first fit the single-segment PSDs. The frequencies of the QPOs detected above $3\sigma$ are shown as underlying gray points in Fig.~\ref{fig:qpo_evolution_mjd}b. We find that the QPO can evolve in frequency on a timescale of a day. There are 62 segments of the length of 500~s between Apr 19 and June 26. We combine successive segments into 10 data epochs. The PSDs, the best-fit models and the QPO profiles are shown in Fig.~\ref{fig:new_qpo}.
The time evolution of the frequencies of the QPOs detected above $3\sigma$ among the 10 epochs is shown by the purple points in Fig.~\ref{fig:qpo_evolution_mjd}. The corresponding information of the NICER observations, the QPO properties including its frequency, Q factor, fractional rms amplitude, and detection significance are summarized in Table~\ref{tab}. All the QPOs detected above $3\sigma$ in multiple-segment PSDs are also detected in all single-segment PSDs, except for the QPOs in Epochs 1 and 2. These two QPOs are also the least significant ones (below $6\sigma$) in Table~\ref{tab}.}

The QPO can be highly coherent as the Q factor is extremely high in Epochs~3--9 ($\gtrsim50$). {\color{black}This is also one of the highest coherence detected for LFQPOs in BHXBs; a LFQPO at 11~mHz has been detected in the hard state of a BHXB H1743--322 with a Q factor as high as $\sim100$ \citep{2012ApJ...754L..23A}.} The QPO amplitude is several percent fractional rms. {\color{black}The QPO frequency is in the range of 5--8~Hz. In Epochs~1--6, the QPO frequency is positively correlated with the count rate (Fig.~\ref{fig:qpo_evolution_mjd}). In Epoch~7, we detect two QPO components in the PSD (Fig.~\ref{fig:new_qpo}g), which are both extremely narrow (Q factor $\gtrsim60$). The lower-frequency component is stronger in terms of the fractional rms amplitude. The QPO frequency ratio is $1.314^{+0.008}_{-0.010}$, indicating that the two components are not in a 2:1 harmonic relationship, which is usually seen for LFQPOs {\color{black}(see Section~\ref{sect:discussion_heartbeat} for the discussion)}. If we identify the lower-frequency component as the fundamental, in Epochs~7--10, the QPO frequency is also positively correlated with the count rate (Fig.~\ref{fig:qpo_evolution_mjd}).}

\begin{table*}
\begin{center}
\caption{The data epochs and the detected highly-coherent QPOs {\color{black}when we fit the PSDs in Fig.~\ref{fig:new_qpo}}. \label{tab}}
\footnotesize
\begin{tabular}{ccccccccc}\hline \hline
Epoch & Obs.IDs & Date & Expo. & Total frac.& QPO Freq. & Q factor & Frac. rms  & Significance  \\ 
& & & (ks) &  rms (\%) & (Hz) & & (\%) & \\
\hline
1&5618010302--5618010403 & 04/19--04/23 & 8.0 & 4.6 & {\color{black}$7.1^{+0.3}_{-0.1}$} & $6^{+6}_{-2}$ & $2.0^{+0.5}_{-0.4}$ & $4.4\sigma$ \\ 
2&5618010404--5618010406 & 04/24--04/26 & 3.5 & 4.5 & $6.69\pm0.09$ & $15^{+8}_{-5}$ & $2.3^{+0.3}_{-0.4}$ & $5.9\sigma$\\ 
3&5618010502--5618010503 & 05/20--05/21 & 3.0 & 7.2 &{\color{black}$6.69\pm0.01$} & $44^{+11}_{-8}$ & $4.1\pm0.2$ & $16.5\sigma$ \\ 
4&5618010504--5618010506 & 05/22--05/24 & 1.0 & 6.8 &{\color{black}$6.81\pm0.02$} & $52^{+16}_{-12}$ & $4.1^{+0.3}_{-0.4}$ & $9.6\sigma$ \\ 
5&5618010507--5618010508 & 05/25--05/26 & 4.5 & 7.2 &{\color{black}$7.07\pm0.01$} & $46^{+5}_{-4}$ & {\color{black}$4.6\pm0.2$} & $22.5\sigma$ \\ 
6&5618010601 & 05/27 & 1.5 &6.3 & {\color{black}$7.22^{+0.01}_{-0.02}$} & $59^{+14}_{-10}$ & $4.2\pm0.3$ & $12.5\sigma$ \\ 
7&5618010602--5618010603 & 05/28--05/29 & 1.5  & 6.1 &
{\color{black}$5.69\pm0.01$} & $63^{+18}_{-16}$ & $3.0\pm0.3$ & $9.8\sigma$ \\ 
& & & & & $7.48^{+0.03}_{-0.04}$ & $58^{+50}_{-26}$ & $1.9\pm0.4$ & $4.8\sigma$ \\ 
8&5618010704--5618010708 & 06/07--06/11 & 4.0 & 5.2 & $6.36\pm0.01$ & $49\pm9$ & $2.8\pm0.2$ & $14.5\sigma$\\ 
9&5618010903--5618010905 & 06/21--06/23 & 2.5 & 4.2 &
{\color{black}$7.10^{+0.02}_{-0.01}$} & $59\pm12$ & $2.4\pm0.2$  & $10.0\sigma$ \\ 
10&5618010907--5618010908 & 06/25--06/26 & 1.5 & 5.0 &
$6.89\pm0.03$ & $31^{+12}_{-8}$ & $2.8\pm0.3$  & $8.4\sigma$ \\ 
\hline
\hline
\end{tabular}
\\
\raggedright{\textbf{Notes.} \\
The total fractional rms is in 0.01--10~Hz. The Q factor is defined as the QPO centroid frequency divided by the width. {\color{black}The fractional rms quoted is the square root of the integrated power density (i.e., for our fits using XSPEC, it is computed as the square root of the normalization of the Lorentzian used to fit the QPO), as we adopt the rms-squared normalization for the PSDs and the normalization of the Lorentzian means the integrated power. The significance of the QPOs is given as the ratio of the integrated power of the Lorentzian used to fit the QPO (i.e., the Lorentzian normalization) divided by the negative $1\sigma$ error on the integrated power. {\color{black}All the uncertainties quoted in this table are for a 90\% confidence range.} 
}}
\end{center}
\end{table*}

\begin{figure}
\centering
\includegraphics[width=1.\linewidth]{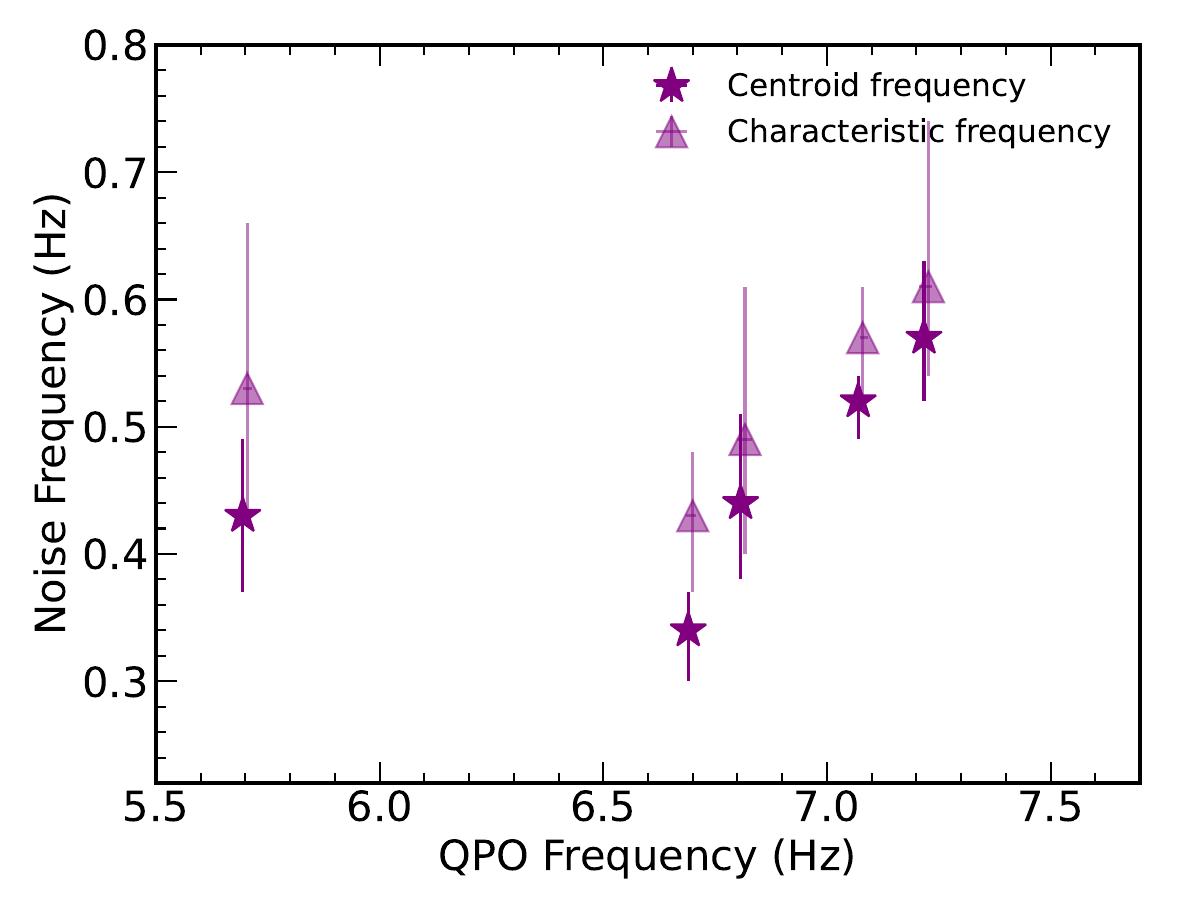}
\caption{{\color{black}The noise frequency vs the QPO frequency in Epochs~3--7 where one noise component with non-zero centroid frequency is required. The stars are for the centroid frequencies, and the triangles are for the characteristic frequencies (see Section~\ref{sect:results} for more details).}}
\label{fig:qpo_freq_noise_freq}
\end{figure}

In addition to the Poisson noise (PSD is flat over frequency), we also observe red noise in all epochs (an empirical description for the power is PSD $\propto f^{-2}$). In addition, a flat-top noise component is required in Epochs 1--2 and 8--10, which can be modeled with a Lorentzian with the centroid frequency fixed at 0, and the {\color{black}FWHM} is in a range of 1--2~Hz. In Epochs~3--7, the centroid frequency of the Lorentzian for the additional noise component is non-zero. It increases from 0.34 to 0.57~Hz through Epochs~3 to 6, as the count rate and the QPO frequency increase. In Epoch~7 where the two QPO components are present simultaneously, the noise centroid frequency decreases to 0.43~Hz. The corresponding Lorentzian {\color{black}FWHM} ranges between 0.4 and 0.6~Hz. {\color{black}It is worth noting that the highly-coherent QPO is the strongest and narrowest when the noise component with the non-zero centroid frequency is present. }{\color{black}We show in Fig.~\ref{fig:qpo_freq_noise_freq} the frequency of this noise component in Epochs~3--7 and the corresponding QPO frequency. In Epochs~3 to 6, the centroid frequency ($\nu_0$) of the noise component is correlated with the QPO frequency; the characteristic frequency ($\sqrt{\nu_0^2+\Delta^2}$ where $\Delta$ is the half width at half maximum) of the noise component is less constrained but the correlation is consistent.} This noise component could be the low-rms extension of the Lorentzian component representing heartbeat-like exotic variability in Class V (see Fig.~\ref{fig:rms_f0}a). {\color{black}Therefore, although the highly-coherent QPO is always present along with characteristics of a soft state including the low total rms and a disk-dominated spectrum, data with the non-zero-centered noise component is classified as Class V rather than the soft state. However, exotic (structured and repeated) variability is too weak to be identified directly in the lightcurves.} 

\begin{figure}
\includegraphics[width=8cm]{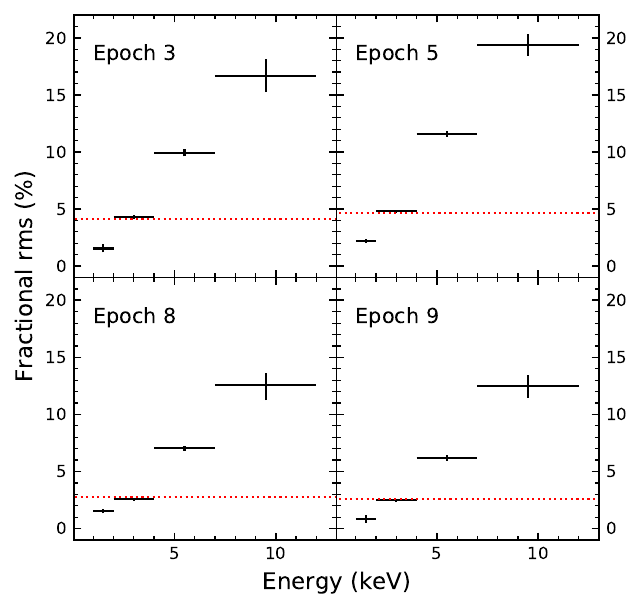}
\caption{{\color{black}Four representative examples of the energy dependence of the QPO's fractional rms. The red dashed line marks the fractional rms in the 1--10 keV band}.}\label{fig:rms_energy}
\end{figure}

{\color{black}We also performed an analysis of the rms-dependence on energy. In addition to the 1-10 keV band, PSDs were created in the 0.5--1 keV, 1--2 keV, 2--4 keV, 4--7 keV, and 7--12 keV bands. All power spectra were rms-normalized. For observations and epochs with the strongest QPOs the PSDs from all energy bands (including 1--10 keV) were fitted simultaneously with a constant and one or two Lorentzians. QPO frequency and width were linked between all power spectra, while the underlying noise level and QPO rms were left to vary. In Figure \ref{fig:rms_energy} we show several representative examples of the rms-dependence on energy. As can be been, the rms has a steep dependence on energy. Note that the QPO was never detected in the 0.5--1 keV band and only upper limits could be determined that were not very constraining; therefore these upper limits are not shown. The QPO typically had a fractional rms values of $\sim$1--2\% in the 1--2 keV band, increasing to $\sim$12--20\% in the 7--12 keV band. }

\begin{figure*}
\centering
\includegraphics[width=1.\linewidth]{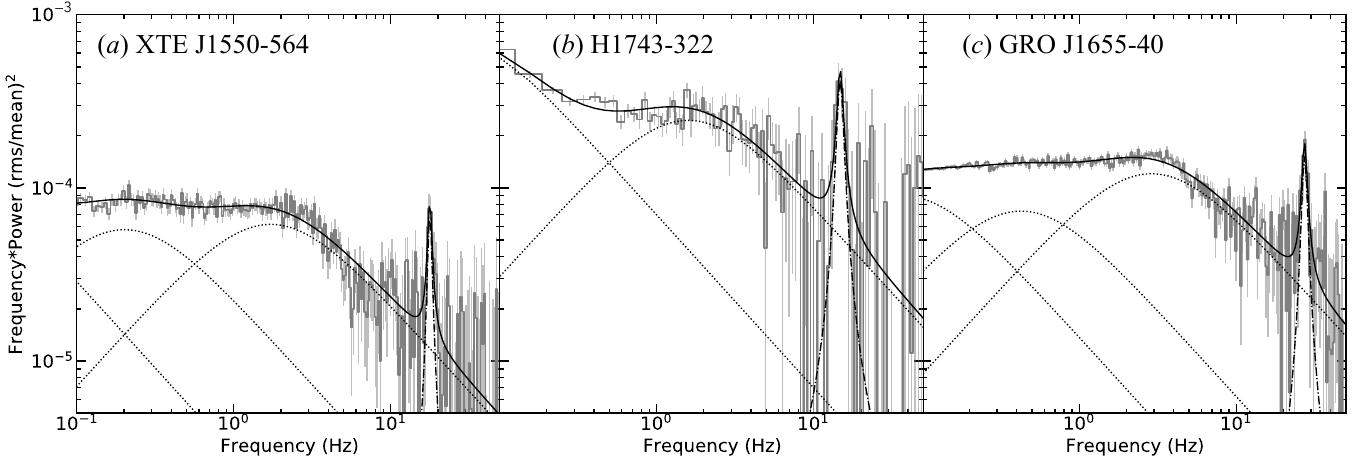}
\caption{Previously detected Type-C QPOs in the soft state of BHXBs using RXTE data including (a) XTE J1550--564 \citep{2001ApJS..132..377H}, (b) H1743--322 \citep{2005ApJ...623..383H}, and (c) GRO J1655--40 \citep{1999ApJ...522..397R,2012MNRAS.427..595M}. {\color{black}The RXTE PSDs are the same as in those references.} The Poisson noise is subtracted to see the QPOs more clearly. The solid line shows the best-fit with multiple Lorentzians, the dotted lines show the noise components, and the dot-dashed lines are the QPOs. The segment length to generate the PSDs is 128~seconds in (a) and (c), and is 16~seconds in (b). The f factor is 0.025. See Table~\ref{tab:other_sources} for the observation information and measurements of the fitted QPOs.}
\label{fig:other_sources}
\end{figure*}

\begin{table*}
\begin{center}
\caption{The QPOs previously detected in the soft state of 3 BHXBs using RXTE data. }\label{tab:other_sources}
\footnotesize
\begin{tabular}{ccccccc}\hline \hline
Source & Energy (keV) & Expo. (ks) & QPO Frequency (Hz) & Q factor & Frac. rms (\%) & Significance\\ 
\hline
XTE J1550--564 & 2--60 & 30.7 & $17.7\pm0.2$ & $15^{+12}_{-6}$ & $0.27\pm0.05$ & $5.1\sigma$\\ 
H1743--322 & 3.7--27.4 & 13.6 & $14.7\pm0.4$ & $10^{+10}_{-5}$ & {\color{black}$0.8\pm0.2$} & $4.0\sigma$\\
GRO J1655--40 & 2--60 & 487.3 & $27.2\pm0.3$ & $11^{+4}_{-3}$ & $0.47\pm0.05$ & $8.3\sigma$\\ 
\hline
\hline
\end{tabular}
\\
\end{center}
\end{table*}

\section{Discussion} \label{sect:discussion}


In the following, we will discuss several possibilities for the nature of the highly coherent QPO we detected in the BHXB \igr\ {\color{black}that can show `heartbeat'-like variability
}.

\subsection{{\color{black}Under the A/B/C classification scheme for LFQPOs}}\label{sect:discussion_typeC}

Conventionally, when the spectrum is dominated by the accretion disk, the {\color{black}QPO classified and the corresponding accretion state} are either Type-B QPO in the SIMS, or Type-C QPO in the soft state.

In the soft state, the PSDs usually contain only a weak (broken) power-law noise and no QPOs. Soft state QPOs have only been detected in a few BHXBs, including XTE~J1550--564 \citep{2001ApJS..132..377H}, H1743--322 \citep{2005ApJ...623..383H}, and GRO~J1655--40 \citep{1999ApJ...522..397R,2012MNRAS.427..595M}. All detections were made in RXTE data. After the detections in these three sources, \citet{2017MNRAS.467..145F} performed a systematic search of QPOs in soft states and found them in three additional sources (GX 339--4, 4U 1543--47, and XTE J1817--330). In Fig.~\ref{fig:other_sources} we show soft state PSDs exhibiting QPOs of the initial three sources in which soft state QPOs were found. Their corresponding observation information and the properties of the QPOs are shown in Table~\ref{tab:other_sources}. The QPOs are at $\sim$15--27~Hz with a Q factor $\sim10$, and their fractional rms amplitudes are $<1$\% (in the {\color{black}hard energy} band; see Table~\ref{tab:other_sources}). These QPOs were classified as Type-C based on the relationship between the noise break frequency and the QPO frequency \citep{2005ApJ...623..383H}, and/or the total fractional rms versus QPO frequency relation \citep{2012MNRAS.427..595M} where the QPO frequency increases and total rms decreases as the source evolves through the outburst.

In Fig.~\ref{fig:rms_f0}b, we show the QPO frequency versus the integrated fractional rms for \igr\ for the highly-coherent QPOs, as well as for the Type-B and Type-C QPOs {\color{black}detected in the same outburst} (see W24a for more details). 
We find that Type-B QPOs lie below the branch traced out by the Type-C QPOs, as found in \citet{2011MNRAS.418.2292M}; the highly coherent QPO could either be a low-rms/high-frequency extension of the Type-C QPOs detected {\color{black}in} this outburst in the hard state and HIMS, or it could be a horizontal extension (with a gap in frequency) of the Type-B QPOs detected in {\color{black}the} SIMS.
In addition, we find that the QPO frequency is correlated with count rate, which is similar to its correlation with the disk flux previously found for Type-C QPOs \citep{2000ApJ...531..537S,remillard2006x,2011MNRAS.418.2292M}. {\color{black} The highly-coherent QPOs in \igr\ show steep energy dependence of their rms, increasing by factors of almost 10 between 1 and 12 keV. This was also observed for the soft state QPOs in XTE J1550-564 \citep{2001ApJS..132..377H}, albeit at higher energies. We note, however, that a steep rms dependence on energy has also been observed for some Type-B QPOs \citep{2001ApJS..132..377H}.} 

{\color{black}If this highly coherent QPO is a Type-C QPO in the soft state, as in the aforementioned 6 BHXBs, the QPO frequency} would correspond to the Lense-Thirring precession frequency at the innermost stable circular orbit (ISCO) radius \citep{2014MNRAS.437.2554M,2014MNRAS.439L..65M}. 
In the solid-body precession model \citep{2009MNRAS.397L.101I}, this means the disk extends very close to the ISCO, leaving only a tiny inner flow `ring' to precess and produce the QPO; {\color{black}it} then becomes unclear how the frame dragging effect could drive such a precession for such a tiny ring of inner flow (see also results and discussion in \citealt{2022MNRAS.511..255N}). In the framework of {\color{black}the} Relativistic Precession Model \citep{1998ApJ...492L..59S}, this means that there is a hot spot or some over-density at the ISCO producing the QPO. However, we note that this is the limiting case of the processing flow when the radial extent of the processing flow is very small \citep{2018MNRAS.473..431M}, so a similar challenge is encountered in this model.

However, there are some discernible differences between the soft-state QPO properties in \igr\ compared to {\color{black}other sources (the three representatives in Fig.~\ref{fig:other_sources} and those in \citealp{2017MNRAS.467..145F})}: (1) the Q factor is much higher ($\gtrsim50$ compared to $\sim10$), (2) the QPO frequency is lower ({\color{black}5--8}~Hz compared to {\color{black}10--27}~Hz), and (3) the QPO is much stronger (fractional rms of the QPO is several percent compared to $<1$\%).

\subsection{A new type of LFQPO in {\color{black}BHXBs that can show `heartbeat'-like variability?}}\label{sect:discussion_heartbeat}

The extremely high coherence of the QPO we detected is in \igr; this BHXB is special in some regards as it is one of the only two BHXBs known to exhibit heartbeat-like variabilities. This suggests that the QPO could also be a new type of LFQPO that is only present in {\color{black}BHXBs that can show `heartbeat'-like variability}, perhaps some sort of resonance is amplified by exotic variability.

\begin{figure}
\includegraphics[width=1.\linewidth]{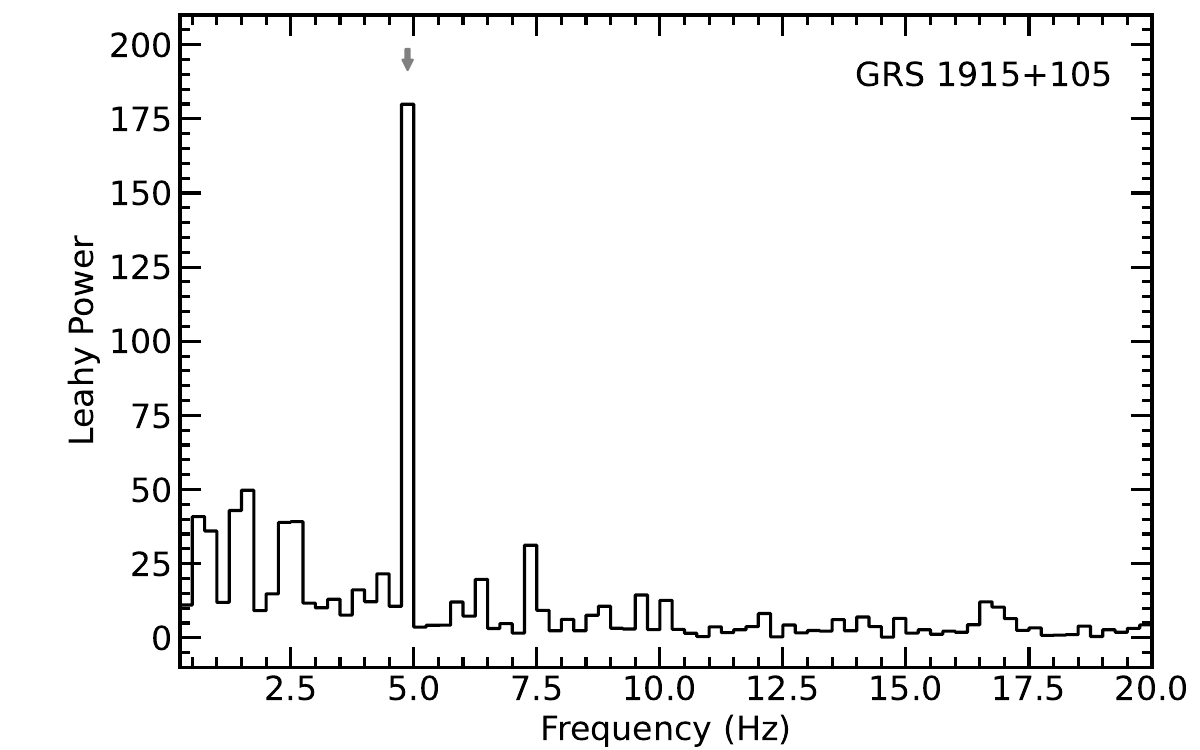}
\caption{{\color{black}Highly-coherent QPO (Q factor $>20$) in \grs. {\color{black}The PSD is computed using two continuous segments of length of 4~seconds, which is a part of the RXTE data 20402-01-45-03 when the lightcurve exhibited heartbeat-like variability (see \citealp{1999ApJ...513L..37M}). The QPO was detected therein, but here, we show it is also highly coherent.} 
The PSD is not rebinned in frequency and is in Leahy normalization. The gray arrow indicates the QPO in the frequency bin 4.75--5~Hz. }}\label{fig:grs_psd}
\end{figure}

The first piece of evidence for this is that in the other {\color{black}source that can show `heartbeat'-like variability}, \grs, the LFQPOs observed along with exotic variabilities (e.g., \citealp{1999ApJ...513L..37M}) can also be highly coherent. One example is shown in Fig.~\ref{fig:grs_psd}, where a QPO detected in the frequency bin 4.75--5~Hz has a Q factor $>20$.
\citet{1999ApJ...513L..37M} also found that above 4~Hz, the QPO frequency depends nearly linearly on the disk flux, consistent with the dependency of the QPO frequency on the count rate found in Section~\ref{sect:results}.

{\color{black}Another piece of information is that} in Section~\ref{sect:results}, we found the highly-coherent QPO is the strongest and narrowest when accompanied by a Lorentzian component peaking at $\sim0.3$--0.6~Hz (Epochs~3--7). Without this Lorentzian component, the QPO properties (the Q factor and fractional rms amplitude) are more in line with those of the Type-C QPOs in the soft state in other sources. This Lorentzian component is likely an extension of the component representing the heartbeat-like exotic variability (see Fig.~\ref{fig:rms_f0}a), even though we cannot identify any exotic (structured and repeated) variability in the light curves. The centroid frequency of the noise component also correlates with the frequency of the highly-coherent QPO (Fig.~\ref{fig:qpo_freq_noise_freq}).



{\color{black}There are two other properties of the highly-coherent QPO worth discussing: the overall PSD shape, and the QPO pair with a frequency ratio inconsistent with 2:1 that is usually observed for LFQPOs.}
The PSDs in Epochs~3--7 (red noise, the noise component peaking at $\sim0.3$--0.6~Hz, and the highly-coherent QPO on top of Poisson noise) are very similar to the ones in which HFQPOs are detected at $\sim67$~Hz in both \igr\ and \grs\ (see Fig.~2 in \citealp{2012ApJ...747L...4A}). The HFQPOs are detected in the variability class $\gamma$ as defined in \citet{2000A&A...355..271B}
, so the low-frequency `bump' is due to exotic variability. {\color{black}In this variability class, there are occasional flares and drops in the lightcurves (see e.g., Fig.~4 in \citealp{2012ApJ...747L...4A}), which are not observed in the data where we detect the highly-coherent QPO.} 

{\color{black}We detect two QPO components in Epoch~7, with a frequency ratio of $1.314^{+0.008}_{-0.010}$.  We tried various selections on count rate and/or hardness ratio, but found the two QPOs are always simultaneously present. In the ObsID-based PSD analysis with a shorter segment length, the QPO pair is detected in four ObsIDs including the two ObsIDs combined in Epoch~7 (Fig.~\ref{fig:obsid_qpo_freq_mjd}; see also Appendix~\ref{appendix:obsid_based} for more details). When both QPO frequencies increase with count rate, the QPO frequency ratio is $1.310\pm0.003$, consistent over the four ObsIDs. This indicates a close relation between the two QPO components.
This frequency ratio is inconsistent with the 2:1 harmonic relationship usually seen for LFQPOs. {\color{black}It is closer to a small integer ratio of 4:3 but still inconsistent with it by $4\sigma$ and $13\sigma$.} Although a frequency ratio of 4:3 has not been observed for either LFQPOs or HFQPOs before, a 3:2 frequency ratio has been observed merely for HFQPOs \citep{2001ApJ...552L..49S,2002ApJ...580.1030R,remillard2006x}.}
A possible mechanism for these harmonic relations {\color{black}(rather than 2:1)} is some sort of resonance in the system (e.g., \citealp{2001A&A...374L..19A}). 
It is also plausible that the two QPOs are not harmonically related, as the frequency ratio is inconsistent with 4:3 by {\color{black}$>3\sigma$}. In this case, the two QPOs could be produced in two different regions with the same physical mechanism as both QPO frequencies are correlated with the count rate. The two regions evolve in e.g., the size, resulting in the QPO frequency evolution. {\color{black}Nevertheless, it is unclear how to sustain a constant frequency ratio when the two QPOs evolve in frequency. }

\section{Summary} \label{summary}

We discovered a highly-coherent QPO in the BHXB \igr\ {\color{black}that can show `heartbeat'-like variability} in its 2022 outburst in NICER. 
Our major findings are as follows: 
\begin{enumerate}

\item{The Q factor reaches $\gtrsim50$, indicating a high coherence of the QPO. {\color{black}This is also {\color{black}one of} the highest coherence of LFQPOs detected in a BHXB.}}

\item{The QPO frequency is in the range of 5--8~Hz, and is positively correlated with the count rate.}

\item{{\color{black}The QPO amplitude is several percent fractional rms in 1--10~keV. The QPO amplitude also increases with energy, reaching $\sim12$--20\% fractional rms in the 7--12~keV band.}} 

\item{{\color{black}The highly-coherent QPO is detected when the spectrum is dominated by the accretion disk emission. Therefore, under the conventional A/B/C classification scheme for LFQPOs, this QPO could either be a Type-C QPO in the soft state or a Type-B QPO in the SIMS. With the total fractional rms vs QPO frequency plot, both possibilities are viable. However, the extremely narrowness of the QPO still hints at a nonidentical origin compared to conventional LFQPOs.}}

\item{This QPO could also be a new type of LFQPO that is only present in sources {\color{black}that can exhibit `heartbeat'-like variability}. The hints we see include, first, highly-coherent LFQPOs are also detected in the other heartbeat source, \grs. Second, the QPO is strongest when accompanied by a noise component that is likely an extension of the component representing the exotic variability; the QPO frequency also correlates with the centroid frequency of this noise component.}

\item{{\color{black}In four observations, we observe two simultaneous QPO components, with a frequency ratio {\color{black}of $\sim1.3$ that is} inconsistent with a normal 2:1 ratio observed for LFQPOs, but closer to a ratio of 4:3. We also find the frequency ratio is constant over the four observations when the QPO frequencies evolve.}}
\end{enumerate}

\vspace{1cm}
We thank Keith Gendreau, Zaven Arzoumanian, and Elizabeth Ferrara for scheduling and performing the NICER observations. 
JW acknowledges support from the NASA~FINESST Graduate Fellowship, under grant 80NSSC22K1596. JW thanks Gibwa Musoke, Megan Masterson, Mason Ng, Navin Sridhar, and Yerong Xu for useful discussions. 
JW, EK, JAG, GM, and ML acknowledge support from NASA~ADAP grant 80NSSC17K0515. AI and DA acknowledge support from the Royal Society. TMB acknowledges the financial contribution from PRIN INAF 2019 n.15.

\appendix
\section{PSD analysis on individual observations} \label{appendix:obsid_based}

{\color{black} As mentioned in Section \ref{obs}, besides a segment-based PSD analysis, we also performed a PSD analysis for individual ObsIDs. For each ObsID we created time-averaged PSDs using segment lengths of 64~s (instead of 500~s), allowing for more ObsIDs to be included in our analysis. Notably, there are four ObsIDs in which a LFQPO pair is detected: 5618010602--5618010605. In the segment-based analysis, only a single QPO pair was detected in Epoch~7 (which combined ObsIDs 5618010602 and 5618010603), and 5618010604 and 5618010605 were excluded owing to their short exposure times. The time evolution of the QPO frequency is shown in the upper panel of Fig.~\ref{fig:obsid_qpo_freq_mjd}. Only QPOs detected with a significance of $>3\sigma$ are shown, with the exception of the $\sim$7.6 Hz QPO in ObsID 5618010605 (which was only detected at 2.5$\sigma$). The data points are color-coded according to the fractional rms amplitude of the QPO.}{\color{black} 

As in the segment-based anaylsis, a clear increase in QPO frequency is observed starting around MJD 59720. When the second (lower) QPO appears, on MJD 59727, it increased in frequency along with the upper QPO, while maintaining a frequency ratio of 1.310$\pm$0.006 (see lower panel of Fig.~\ref{fig:obsid_qpo_freq_mjd}. While this ratio is close to 4:3, it is not consistent with that value. The upper QPO showed a sudden drop in rms when the lower peak appeared, from $\sim$3.5--4.7 \% to $\sim$1.7--2.9. After MJD 59731 the upper QPO was no longer detected, while the lower QPO continued to increase in frequency until MJD $\sim$59750. The fact that 1) the two QPOs had a constant frequency ratio (albeit in a narrow {\color{black}frequency} range) and 2) the upper QPO became significantly weaker when the lower QPO appeared, suggests a connection between the mechanisms responsible for the two QPOs.}

\setcounter{figure}{0}
\renewcommand{\thefigure}{A\arabic{figure}}
\renewcommand\theHfigure{Appendix.\thefigure}

\setcounter{table}{0}
\renewcommand{\thetable}{A\arabic{table}}

\begin{figure}
\centering
\includegraphics[width=8.5cm]{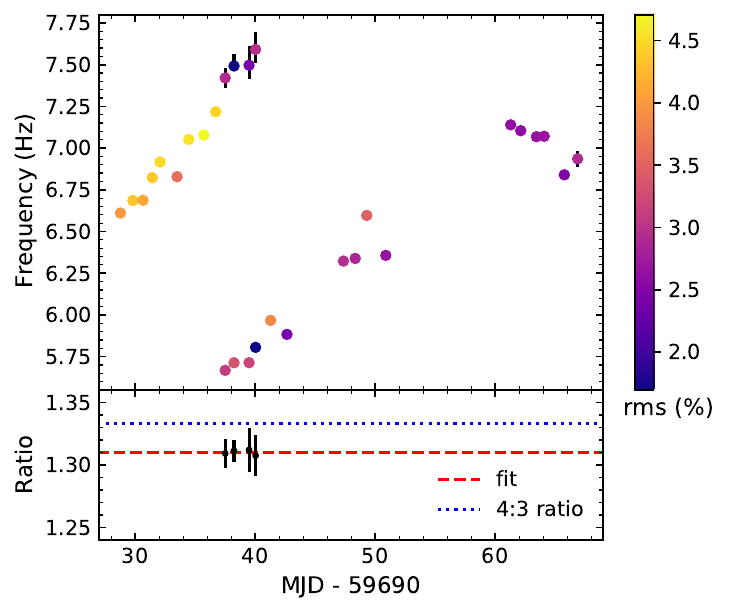}
\caption{{\color{black}The time evolution of the QPO frequency (upper panel) and QPO frequency ratio (lower panel, when applicable) for individual NICER ObsIDs. Data points in the upper panel are color-coded according the their fractional rms value (for rms scale, see color bar on the right side). The frequency error uncertainties are generally smaller than the symbol size. The best fit with a constant to the frequency ratios is depicted by the dashed red line (value of 1.310); for reference we also show the a 4:3 ratio with a dotted blue line.  }}
\label{fig:obsid_qpo_freq_mjd}
\end{figure}

\bibliographystyle{apj}
\bibliography{draftv5_arXiv}
\end{document}